\documentstyle[amssymb,psfig]{mn}

\newif\ifAMStwofonts
\ifoldfss

  \ifCUPmtlplainloaded \else
    \NewTextAlphabet{textbfit} {cmbxti10} {}
    \NewTextAlphabet{textbfss} {cmssbx10} {}
    \NewMathAlphabet{mathbfit} {cmbxti10} {} 
    \NewMathAlphabet{mathbfss} {cmssbx10} {} 
  \fi
  
\ifAMStwofonts
  
\ifCUPmtlplainloaded \else
      \NewSymbolFont{upmath} {eurm10}
      \NewSymbolFont{AMSa} {msam10}
      \NewMathSymbol{\upi}     {0}{upmath}{19}
      \NewMathSymbol{\umu}     {0}{upmath}{16}
      \NewMathSymbol{\upartial}{0}{upmath}{40}
      \NewMathSymbol{\leqslant}{3}{AMSa}{36}
      \NewMathSymbol{\geqslant}{3}{AMSa}{3E}

    \fi
    
\fi
\fi 
\ifnfssone
  
\newmathalphabet{\mathit}
  \addtoversion{normal}{\mathit}{cmr}{m}{it}
  \addtoversion{bold}{\mathit}{cmr}{bx}{it}

  \newmathalphabet{\mathbfit} 
  \addtoversion{normal}{\mathbfit}{cmr}{bx}{it}
  \addtoversion{bold}{\mathbfit}{cmr}{bx}{it}
  
\newmathalphabet{\mathbfss} 
  \addtoversion{normal}{\mathbfss}{cmss}{bx}{n}
  \addtoversion{bold}{\mathbfss}{cmss}{bx}{n}
  
\ifAMStwofonts
    
\ifCUPmtlplainloaded \else
      
      \UseAMStwoboldmath
      
\makeatletter
      \new@mathgroup\upmath@group
      \define@mathgroup\mv@normal\upmath@group{eur}{m}{n}
      \define@mathgroup\mv@bold\upmath@group{eur}{b}{n}
      \edef\UPM{\hexnumber\upmath@group}
      
\new@mathgroup\amsa@group
      \define@mathgroup\mv@normal\amsa@group{msa}{m}{n}
      \define@mathgroup\mv@bold\amsa@group{msa}{m}{n}
      \edef\AMSa{\hexnumber\amsa@group}
      \makeatother
      
\mathchardef\upi="0\UPM19
      \mathchardef\umu="0\UPM16
      \mathchardef\upartial="0\UPM40
      \mathchardef\leqslant="3\AMSa36
      \mathchardef\geqslant="3\AMSa3E

    \fi
  \fi
  
\fi 
\ifnfsstwo

  \DeclareMathAlphabet{\mathbfit}{OT1}{cmr}{bx}{it}
  \SetMathAlphabet\mathbfit{bold}{OT1}{cmr}{bx}{it}
  \DeclareMathAlphabet{\mathbfss}{OT1}{cmss}{bx}{n}
  \SetMathAlphabet\mathbfss{bold}{OT1}{cmss}{bx}{n}
  
\ifAMStwofonts
    
\ifCUPmtlplainloaded \else
      \DeclareSymbolFont{UPM}{U}{eur}{m}{n}
      \SetSymbolFont{UPM}{bold}{U}{eur}{b}{n}
      \DeclareSymbolFont{AMSa}{U}{msa}{m}{n}
      \DeclareMathSymbol{\upi}{0}{UPM}{"19}
      \DeclareMathSymbol{\umu}{0}{UPM}{"16}
      \DeclareMathSymbol{\upartial}{0}{UPM}{"40}
      \DeclareMathSymbol{\leqslant}{3}{AMSa}{"36}
      \DeclareMathSymbol{\geqslant}{3}{AMSa}{"3E}

    \fi
  \fi
  
\fi 
\ifCUPmtlplainloaded \else
  \ifAMStwofonts \else 
    
\def\upi{\pi}
\def\umu{\mu}
\def\upartial{\partial}
  \fi
\fi
\pubyear{1994}

\title{Small-scale microwave background anisotropies due to tangled primordial
magnetic fields}
\author[Kandaswamy Subramanian and John. D. Barrow]
{Kandaswamy Subramanian$^{1}$, and John. D. Barrow$^2$ \\
$^{1}$Inter University Centre for Astronomy and Astrophysics, \\
Post Bag 4, Ganeshkhind, Pune 411 007, India\\
$^{2}$DAMTP, Centre for Mathematical Sciences, Cambridge University,
Wilberforce Road, \\
Cambridge CB3 0WA, UK}
\date{}
\begin{document}
\maketitle

\begin{abstract}
An inhomogeneous cosmological magnetic field creates vortical perturbations
that survive Silk damping on much smaller scales than compressional modes.
This ensures that there is no sharp cut-off in anisotropy on arc-minute
scales. As we had pointed out earlier, tangled magnetic fields, if they
exist, will then be a potentially important contributor to small-angular
scale CMBR anisotropies. Several ongoing and new experiments, are expected
to probe the very small angular scales, corresponding to multipoles with $%
l>1000$. In view of this observational focus, we revisit the predicted
signals due to primordial tangled magnetic fields, for different spectra and
different cosmological parameters. We also identify a new regime, where the
photon mean-free path exceeds the scale of the perturbation, which dominates
the predicted signal at very high $l$. A scale-invariant spectrum of tangled
fields which redshifts to a present value $B_{0}=3\times 10^{-9}$ Gauss,
produces temperature anisotropies at the $10\mu K$ level between $l\sim
1000-3000$. Larger signals result if the univese is lambda dominated, if the
baryon density is larger, or if the spectral index of magnetic tangles is
steeper, $n>-3$. The signal will also have non-Gaussian statistics. We
predict the distinctive form of the increased power expected in the
microwave background at high $l$ in the presence of significant tangled
magnetic fields. We may be on the verge of detecting or ruling out the
presence of tangled magnetic fields which are strong enough to
influence the formation of large-scale structure in the Universe.
\end{abstract}

\label{firstpage}%

\section{Introduction}

The origin of large-scale cosmic magnetic fields remains a challenging
problem. It is widely assumed that magnetic fields in astronomical objects,
like galaxies, grew by turbulent dynamo action on small seed magnetic fields
(cf. Ruzmaikin, Shukurov \& Sokoloff 1988; Beck et al 1996). However, the
efficiency of turbulent galactic dynamos is still unclear, especially in
view of the constraints implied by helicity conservation (Cattaneo \&
Vainshtein 1991; Kulsrud \& Anderson 1992; Gruzinov \& Diamond 1994;
Cattaneo \& Hughes 1996;Subramanian 1998, 1999; Blackman \& Field 2000;
Kleoorin et al 2000; Brandenburg 2001; Brandenburg and Subramanian 2000;
Brandenburg, Dobler and Subramanian 2002). Magnetic fields with larger
coherence scales may also be present in clusters of galaxies (Clarke,
Kronberg \& Bohringer 2001) and at high redshifts (Oren \& Wolfe 1995). It
is not yet clear that all these large-scale coherent fields could result
from dynamo action. Alternatively, galactic or cluster fields could be
remnants of a primordial cosmological magnetic field (cf. Kulsrud 1990;
1999), although, as yet, there is no entirely compelling mechanism for
producing the required field. It could be present in the initial conditions,
be produced quantum gravitationally or at a phase transition, or be
generated in some way at the end of a period of inflation, perhaps with an
almost scale-invariant spectrum (Turner \& Widrow 1988; Ratra 1992; cf.
Grasso and Rubenstein 2001 for a review).

A primordial field that expanded to contribute a present field strength of
order $10^{-9}$ Gauss, tangled on galactic scales, could also affect the
process of galaxy formation (Rees \& Reinhardt 1972; Wasserman 1978; Kim,
Olinto \& Rosner 1996; Subramanian \& Barrow 1998a, SB98 hereafter). It is
of considerable interest, therefore, to find different ways of limiting or
detecting such primordial fields (see Kronberg 1994 and Grasso \& Rubenstein
2001 for reviews). Here, we will show that the imminent extension of cosmic
microwave background radiation (CMBR) observations to very small angular
scales allows us to test for the presence of dynamically significant
cosmological magnetic fields in new ways. This is because tangled magnetic
fields create small-scale power in CMBR temperature anisotropy that would
have been damped out if magnetic fields were absent. Conversely, if enhanced
power is found in the CMBR power spectrum on very small scales then it might
be explained by the effects of tangled magnetic fields.

In an earlier paper (Subramanian \& Barrow 1998b, hereafter Paper I), we
argued that observations of anisotropies in the CMBR provide a potentially
powerful constraint on such tangled magnetic fields. Indeed, the isotropy of
the CMBR already places a limit of $6.8\times 10^{-9}(\Omega _{m}h^{2})^{%
\frac{1}{2}}$ Gauss on the present strength of any {\it uniform} (spatially
homogeneous) component of the magnetic field (Barrow, Ferreira \& Silk
1997), where $\Omega _{m}$ is the present matter density parameter, and $h$
the Hubble constant in units of $100$ km s$^{-1}$ Mpc$^{-1}$). In Paper I we
obtained comparable constraints on tangled (inhomogeneous) magnetic fields
and highlighted the distinctive fluctuation signature that they are expected
to leave in the small-scale structure of the CMBR. In particular, tangled
magnetic fields produce vortical perturbations, which are overdamped in the
radiation era. These can then survive Silk damping (Silk 1968) on scales much smaller
than the compressional modes (Jedamzik, Katalinic \& Olinto 1998; SB98), so
an obvious place to expect signals induced by tangled magnetic fields is
below the Silk damping scale, or at multipoles of greater than $l\sim 1000$.
Several new and ongoing experiments (MAP, VSA, ACBAR, CBI, ATCA and
Planck Surveyor) are indeed expected to provide information in this large $l$
regime. This motivates us to revisit the computation of the expected signals
at larger values of $l$, for a wider variety of cosmological parameters and
spectral indices, than were made in Paper I. In particular, we calculate
results for a new high-$l$ regime, where the photon mean-free path exceeds
the scale of the perturbation. These predictions will allow comparison with
future observations.

\section{ Estimates of the induced anisotropy}

The evolution of temperature anisotropy for vector perturbations has been
derived in detail by Hu and White (1997a) in the total angular momentum
representation and was also given in Paper I. The temperature anisotropy is
expanded in terms of tensor spherical harmonics and the angular power
spectrum induced by vector perturbations is given by (see Paper I and
Mack, Kahniashvilli and Kosowsky 2002), 
\begin{eqnarray}
C_{l} &=&4\pi \int_{0}^{\infty }{\frac{k^{2}dk}{2\pi ^{2}}}\quad {\frac{%
l(l+1)}{2}}  \nonumber \\
\  &&\times <|\int_{0}^{\tau _{0}}d\tau g(\tau _{0},\tau )v(k,\tau ){\frac{%
j_{l}(k(\tau _{0}-\tau ))}{k(\tau _{0}-\tau )}}|^{2}>.  \label{deldef}
\end{eqnarray}
Here, $v(k,\tau )$ is the magnitude of the vorticity, $\Omega
_{i}=v_{i}^{B}-V_{i}$, in Fourier space, where $v_{i}^{B}$ is the rotational
component of the fluid velocity, and $V_{i}$ is the vector metric
perturbation. Note that since $\Omega _{i}$ also appears in the Euler
equation for baryons, we need never compute the vector metric perturbation
explicitly (see also Mack, Kahniashvilli and Kosowsky 2002). Also, $k$ is
the co-moving wave number, $\tau $ is conformal time, $\tau _{0}$ its
present value, and $j_{l}(z)$ is the spherical Bessel function of order $l$.
We have ignored a small polarization correction to the source term
and also a metric perturbation term which is subdominant at
large $l$ (cf. Seshadri and Subramanian 2001). The
'visibility function', $g(\tau _{0},\tau ),$ determines the probability that
a photon reaches us at epoch $\tau _{0}$ if it was last scattered at the
epoch $\tau $. We adopt a flat universe throughout, with a total matter
density $\Omega _{m}$ and a non-zero cosmological constant density $\Omega
_{\Lambda }=1-\Omega _{m}$ today.

In Paper I we obtained analytic estimates for $C_{l}$ in various asymptotic
regimes. We briefly recapitulate the arguments and results. Firstly, we
approximated the visibility function as a Gaussian: $g(\tau _{0},\tau
)=(2\pi \sigma ^{2})^{-1/2}\exp [-(\tau -\tau _{\ast })^{2}/(2\sigma ^{2})]$%
, where $\tau _{\ast }$ is the conformal epoch of ``last scattering'' and $%
\sigma $ measures the width of the LSS. To estimate these, we use the
expressions given in Hu and Sugiyama (1995), adopting a baryon density
parameter $\Omega_{b}=0.02h^{-2}$, and $h=0.7$. We estimate the epoch of
last scattering, to be $(1+z_{\ast })\sim 1150$ and $\sigma /\tau _{\ast
}\sim 0.064$. To convert redshift into conformal time we use $\tau
=6000h^{-1}((a+a_{eq})^{1/2}-a_{eq}^{1/2})/\Omega _{m}^{1/2}$, valid for a
flat universe (cf. Hu \& White 1997b). Here, the expansion factor $%
a=(1+z)^{-1}$ and $a_{eq}=4.17\times 10^{-5}(\Omega_{m}h^{2})^{-1}$. For an $%
\Omega _{m}=1$ model, we get $\tau _{\ast }=130.0h^{-1}$ Mpc, and $\sigma
=8.4h^{-1}$ Mpc, while for a $\Lambda $-dominated model with $\Omega _{m}=0.3
$, we get $\tau _{\ast }=187.5h^{-1}$ Mpc and $\sigma =12.1h^{-1}$ Mpc. We
will use these numbers in the numerical estimates below.

The dominant contributions to the integral over $\tau $ in Eq. (\ref{deldef}%
) then come from a range $\sigma $ around the epoch $\tau =\tau _{\ast }$.
Furthermore, $j_{l}(k(\tau _{0}-\tau ))$ picks out $(k,\tau )$ values in the
integrand which have $k(\tau _{0}-\tau )\sim l.$ Thus, following the
arguments detailed in Paper I, for $k\sigma <<1$ we get the analytical
estimate, $l(l+1)C_{l}/(2\pi )\approx (\pi /4)\Delta _{v}^{2}(k,\tau _{\ast
})|_{k=l/R_{\ast }}$. Here, $\Delta _{v}^{2}=k^{3}<|v(k,\tau _{\ast
})|^{2}>/(2\pi ^{2})$ is the power per unit logarithmic interval of $k$,
residing in the {\it net} vorticity perturbation, and $R_{\ast }=\tau
_{0}-\tau _{\ast }$. In the opposite limit, $k\sigma >>1$, we get $%
l(l+1)C_{l}/(2\pi )\approx (\sqrt{\pi }/4)(\Delta _{v}^{2}(k,\tau _{\ast
})/(k\sigma )|_{k=l/R_{\ast }}$. At small wavelengths, $C_{l}$ is suppressed
by a $1/k\sigma $ factor due to the finite thickness of the LSS.

To evaluate $C_{l}$, one also needs to estimate $v$, the vorticity induced
by magnetic inhomogeneities. We assume the magnetic field to be initially a
Gaussian random field. On galactic scales and above, the induced velocity is
generally so small that it does not lead to any appreciable distortion of
the initial field (Jedamzik, Katalinic and Olinto 1998, SB98). So, to a very
good approximation, the magnetic field simply redshifts away as ${\bf B}(%
{\bf x},t)={\bf b}_{0}({\bf x})/a^{2}$. The Lorentz force associated with
the tangled field is then ${\bf F}_{L}=({\bf \nabla }\times {\bf b}%
_{0})\times {\bf b}_{0}/(4\pi a^{5})$, which pushes the fluid and creates
rotational velocity perturbations. These can be estimated as in Paper I, by
using the Euler equation for the baryons. On scales larger than the photon
mean-free-path at decoupling, where the viscous effect due to photons can be
treated in the diffusion approximation, this reads 
\begin{equation}
\left( {\frac{4}{3}}\rho _{\gamma }+\rho _{b}\right) {\frac{\partial \Omega
_{i}}{\partial t}}+\left[ {\frac{\rho _{b}}{a}}{\frac{da}{dt}}+{\frac{%
k^{2}\eta }{a^{2}}}\right] \Omega _{i}={\frac{P_{ij}F_{j}}{4\pi a^{5}}}.
\label{eulerk}
\end{equation}
Here, $\rho _{\gamma }$ is the photon density, $\rho _{b}$ the baryon
density, and $\eta =(4/15)\rho _{\gamma }l_{\gamma }$ the shear viscosity
coefficient associated with the damping due to photons, whose mean-free-path
is $l_{\gamma }=(n_{e}\sigma _{T})^{-1}\equiv L_{\gamma }a(t)$, where $n_{e}$
is the electron density and $\sigma _{T}$ the Thomson cross-section.
For $(1 + z_\ast) \sim 1150$, we get $L_\gamma(\tau_*) \sim 1.8 f_b^{-1}$ Mpc,
where $f_{b}=(\Omega_{b}h^{2}/0.02)$. We have
defined the Fourier transforms of the magnetic field, by ${\bf b}_{0}({\bf x}%
)=\sum_{{\bf k}}{\bf b}({\bf k})\exp (i{\bf k}.{\bf x})$ and ${\bf F}({\bf k}%
)=\sum_{{\bf p}}[{\bf b}({\bf k}+{\bf p}).{\bf b}^{\ast }({\bf p})]{\bf p}-[%
{\bf k}.{\bf b}^{\ast }({\bf p})]{\bf b}({\bf k}+{\bf p})$. The projection tensor, 
$P_{ij}({\bf k})=[\delta _{ij}-k_{i}k_{j}/k^{2}]$ 
projects ${\bf F}$ onto its transverse components perpendicular to ${\bf k}$.

The comoving Silk damping scale at recombination, $L_{S}=k_{S}^{-1}\sim 10$
Mpc, separates scales on which the radiative viscosity is important ($%
kL_{S}\gg 1$) from those on which it is negligible ($kL_{S}\ll 1$). For $%
kL_{s}\ll 1$, the damping due to the photon viscosity can be neglected
compared to the Lorentz force. Integrating the baryon Euler equation,
assuming negligible initial vorticity perturbation, then gives $\Omega
_{i}=G_{i}\tau /(1+S_{\ast })$, where $G_{i}=3P_{ij}F_{j}/[16\pi \rho _{0}]$%
, $\rho _{0}$ is the present-day value of $\rho _{\gamma }$, and $S_{\ast
}=(3\rho _{b}/4\rho _{\gamma })(\tau _{\ast })\sim 0.53(\Omega
_{b}h^{2}/0.02)$. In the other limit, with $kL_{s}>>1$, we can use the
terminal-velocity approximation, neglecting the inertial terms in the Euler
equation, to balance the Lorentz force by friction. This gives $\Omega
_{i}=(G_{i}/k)(5/kL_{\gamma })$, on scales where diffusion damping operates.
The transition Silk scale can also be estimated by equating $\Omega _{i}$ in
the two cases, to give $k_{S}\sim \lbrack 5(1+S_{\ast })/(\tau L_{
\gamma }(\tau))]^{1/2}$.

A new regime arises on very small scales that are well below the photon
mean-free path. The radiative drag force is then no longer described by the
diffusion approximation, but rather by the free-streaming drag given in Eq.
(6.1) of SB98. Under strong damping, this term dominates the inertial terms
in the Euler equation and the fluid reaches terminal velocity where friction
due to free-streaming photons balances driving by the Lorentz force, ${\bf F}%
_{L}$. From Eq. (6.2) of SB98, we get $4\rho _{\gamma }/(3\rho
_{b})n_{e}\sigma _{T}{\bf \Omega }={\bf F}_{L}/\rho _{b}$, which gives $%
\Omega _{i}=G_{i}L_{\gamma }$. One can also estimate the transition scale,
say $k_{fs}^{-1}$, below which free-streaming damping dominates, by equating
the induced velocities in the diffusion and free-streaming damping regimes,
to get $k_{fs}\sim \sqrt{5}/L_{\gamma }$.

In order to compute the $C_{l}$s we also need to specify the spectrum of the
tangled magnetic field, say $M(k)$. We define, $<b_{i}({\bf k})b_{j}({\bf q}%
)>=\delta _{{\bf k},{\bf q}}P_{ij}({\bf k})M(k)$, where $\delta _{{\bf k},%
{\bf q}}$ is the Kronecker delta which is non-zero only for ${\bf k}={\bf q}$%
. This gives $<{\bf b}_{0}^{2}>=2\int (dk/k)\Delta _{b}^{2}(k)$, where $%
\Delta _{b}^{2}(k)=k^{3}M(k)/(2\pi ^{2})$ is the power per logarithmic
interval in $k$ space residing in magnetic tangles, and we replace the
summation over $k$ space by an integration. The ensemble average $<|v|^{2}>$%
, and hence the $C_{l}$s, can be computed in terms of the magnetic spectrum $%
M(k)$. It is convenient to define a dimensionless spectrum, $h(k)=\Delta
_{b}^{2}(k)/(B_{0}^{2}/2)$, where $B_{0}$ is a fiducial constant magnetic
field. The Alfv\'{e}n velocity, $V_{A}$, for this fiducial field is, 
\begin{equation}
V_{A}={\frac{B_{0}}{(16\pi \rho _{0}/3)^{1/2}}}\approx 3.8\times
10^{-4}B_{-9},  \label{alfvel}
\end{equation}
where $B_{-9}\ \equiv (B_{0}/10^{-9}{\rm Gauss})$. We will also consider
power-law magnetic spectra, $M(k)=Ak^{n}$ cut-off at $k=k_{c}$, where $k_{c}$
is the Alfv\'{e}n-wave damping length-scale (Jedamzik, Katalinic and Olinto,
SB98). We fix $A$ by demanding that the smoothed field strength over a
''galactic'' scale, $k_{G}=1h{\rm Mpc}^{-1}$, (using a sharp $k$-space
filter) is $B_{0}$, giving a dimensionless spectrum for $n>-3$ of 
\begin{equation}
h(k)=(n+3)(k/k_{G})^{3+n}.  \label{powspec}
\end{equation}

We can now put together the above results to derive analytic estimates for
the CMBR anisotropy induced by tangled magnetic fields. As a measure of the
anisotropy we define the quantity $\Delta T(l)\equiv \lbrack
l(l+1)C_{l}/2\pi ]^{1/2}T_{0}$, where $T_{0}=2.728$ K is the CMBR
temperature. On large scales, such that $kL_{s}<1$ and $k\sigma <1$, the
resulting CMBR anisotropy is (see Paper I) 
\begin{eqnarray}
\Delta T_{B}(l) &=&T_{0}({\frac{\pi }{32}})^{1/2}I(k){\frac{kV_{A}^{2}\tau
_{\ast }}{(1+S_{\ast })}}  \nonumber \\
\  &\approx &5.8\mu K\left( {\frac{B_{-9}}{3}}\right) ^{2}\left( {\frac{l}{%
500}}\right) I({\frac{l}{R_{\ast }}}).  \label{largT}
\end{eqnarray}
Here, $l=kR_{\ast }$ and we have used cosmological parameters for the $%
\Lambda $-dominated model, with $\Omega _{\Lambda }=0.7$, $\Omega _{m}=0.3$
and $\Omega_{b}h^{2}=0.02$ (in Paper I, we used a purely matter-dominated $%
\Omega _{m}=1$ model). We also use the fit given by Hu and White (1997b) to
calculate $\tau
_{0}=6000h^{-1}((1+a_{eq})^{1/2}-a_{eq}^{1/2})(1-0.0841ln(\Omega
_{m}))/\Omega _{m}^{1/2}$, valid for flat universe.

On scales where $kL_{S}>1$ and $k\sigma >1$, but $kL_{\gamma }(\tau _{\ast
})<1$, we get 
\begin{eqnarray}
\Delta T_{B}(l) &=&T_{0}{\frac{\pi ^{1/4}}{\sqrt{32}}}I(k){\frac{5V_{A}^{2}}{%
kL_{\gamma }(\tau _{\ast })(k\sigma )^{1/2}}}  \nonumber \\
\  &\approx &13.0\mu K\left( {\frac{B_{-9}}{3}}\right) ^{2}\left( {\frac{l}{%
2000}}\right) ^{-3/2}f_{b}h_{70}^{-1}I({\frac{l}{R_{\ast }}}),  \label{smalT}
\end{eqnarray}
where $h_{70}\equiv (h/0.7)$ and $f_{b}=(\Omega_{b}h^{2}/0.02)$; there is
also a weaker dependence of other parameters on $f_{b}$ but the strongest
dependence comes from the fact that $L_{\gamma }\propto f_{b}^{-1}$. In the
free-streaming regime, the estimate for $\Delta T(l)$ is obtained by
replacing $(kL_{\gamma }/5)^{-1}$ in Eq. (\ref{smalT}) by $(kL_{\gamma })$.
The CMBR anisotropy which results for scales so small that free-streaming
damping dominates is 
\begin{eqnarray}
\Delta T_{B}(l) &=&T_{0}{\frac{\pi ^{1/4}}{\sqrt{32}}}I(k){\frac{%
V_{A}^{2}kL_{\gamma }(\tau _{\ast })}{(k\sigma )^{1/2}}}  \nonumber \\
\  &\approx &0.4\mu K\left( {\frac{B_{-9}}{3}}\right) ^{2}\left( {\frac{l}{%
20000}}\right) ^{1/2}I({\frac{l}{R_{\ast }}}){\frac{h}{f_{b}}}.
\label{versmalT}
\end{eqnarray}

The function $I^{2}(k)$ in the Eqs.(\ref{largT})-(\ref{versmalT}) is a
dimensionless mode-coupling integral given by 
\begin{eqnarray}
I^{2}(k) &=&\int_{0}^{\infty }{\frac{dq}{q}}\int_{-1}^{1}d\mu {\frac{h(q)h(|(%
{\bf k}+{\bf q})|)k^{3}}{(k^{2}+q^{2}+2kq\mu )^{3/2}}}  \nonumber \\
&&\ \times (1-\mu ^{2})\left[ 1+{\frac{(k+2q\mu )(k+q\mu )}{%
(k^{2}+q^{2}+2kq\mu )}}\right] ,  \label{modint}
\end{eqnarray}
where $|({\bf k}+{\bf q}|=(k^{2}+q^{2}+2kq\mu )^{1/2}$. In general, $I(k)$
can only be evaluated numerically but for $h(k)=k\delta _{D}(k-k_{0})$,
where $\delta _{D}(x)$ is the Dirac delta function, it can be evaluated
exactly. One gets, $I(k)=(k/k_{0})[1-(k/2k_{0})^{2}]^{1/2}$, for $k<2k_{0}$,
and zero for larger $k$. So in this case, $I(k)$ contributes a factor of
order unity around $k\sim k_{0}$, with $I(k_{0})=\sqrt{3}/2$.

We can also find an analytic approximation to the mode coupling integral for
power-law magnetic spectra. The approximation is different for $n>-3/2$, and
for $n<-3/2$. For $n>-3/2$ and for $k<<k_{c}$ (which is relevant for $%
l<<k_{c}R_{\ast }$), one gets (Seshadri \& Subramanian 2001): 
\begin{equation}
I^{2}(k)={\frac{28}{15}}{\frac{(n+3)^{2}}{(3+2n)}}({\frac{k}{k_{G}}})^{3}({%
\frac{k_{c}}{k_{G}}})^{3+2n}.  \label{mklar}
\end{equation}
The mode-coupling integral is dominated by the small scale cut-off in this
case. In the other limit $n<-3/2$, we find: 
\begin{equation}
I^{2}(k)={\frac{8}{3}}(n+3)({\frac{k}{k_{G}}})^{6+2n}  \label{mksmal}
\end{equation}
For a nearly scale invariant spectrum, say with $n=-2.9$, we then get $%
\Delta T(l)\sim 4.7\mu K(l/1000)^{1.1}$ for scales larger than the Silk
scale, and $\Delta T(l)\sim 5.6\mu K(l/2000)^{-1.4}$, for scales smaller
than $L_{S}$ but larger than $L_{\gamma }$. Larger signals will be expected
for steeper spectra, $n>-2.9$ at the higher $l$ end.

To complement these analytic results, we have also computed $\Delta T(l)$
for the above spectra, by evaluating the $\tau $ and $k$ integrals in Eq.(%
\ref{deldef}) numerically. We retain the analytic approximations to $I(k)$
and $\Omega _{i}(k),$ with transitions between the limiting forms at
wavenumbers $k_{S}$ and $k_{fs}$. The results are displayed in Figures 1 and
2. We see that for $B_{0}\sim 3\times 10^{-9}G$, this leads to a predicted
RMS temperature anisotropy in the CMBR of order $10\mu K$ for $1000<l<3000$,
for a nearly scale-invariant power law spectra with $n=-2.9$. Larger signals
result, at the high $l$ end, for a $\Lambda $-dominated universe, compared 
to a matter dominated model (compare the solid and dotted curves), 
basically due to an increase in $R_{\ast }$ for this model. Larger signals 
also result for larger baryon density (compare the solid and dashed-triple-dotted curves), 
due to the decrease in $L_{\gamma }$ and hence the damping effects of 
radiative viscosity. Also, a moderately steeper spectral index, with $n=-2.5$, 
which has more power on small scales leads to an increased $\Delta T$ 
(compare the dashed-triple-dotted and dashed curves). Much larger signals result from 
even steeper spectra, but we have not displayed these results, as spectra 
with $n>-2.5$ and $B_{-9}\sim 3$, are probably ruled out because of gravitational 
wave production, estimated by Caprini and Durrer (2002). They will also be 
significantly constrained by the high-$l$ limits on the anisotropy by ATCA, 
given in Subrahmanyan et al (1998). We find that our analytic approximation 
of the $\tau $ and $k$ integrals in Eq.(\ref{deldef}), tends to underestimate 
the amplitude of $\Delta T(l)$ by about a factor $\sim 2$, although the 
analytically predicted $l$-dependences agree very well with the numerical 
integration at both small and large $l$. The numerical integration over 
the Bessel function and the visibility functions are of course expected to be 
more accurate. The new regime of free-streaming damping is only seen at 
very large $l\sim k_{fs}R_{\ast }$, which is of order $10^{4}$ even for 
the matter-dominated model, and is $\sim 20,000$ for the $\Lambda $-dominated model.

\begin{figure}
\begin{picture}(240,240)
\psfig{figure=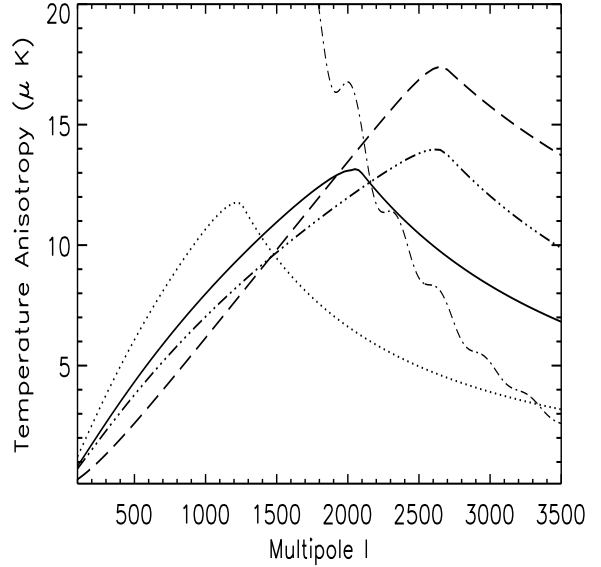,height=8.5cm,width=8.5cm,angle=0.0}
\end{picture}
\caption{$\Delta T$ versus $l$ predictions for different cosmological models
and magnetic power spectrum $M(k)\varpropto k^{n}$, for $B_{-9}=3$. The bold
solid line ({\bf ---}){\bf \ }is for a canonical flat,  $\Lambda $-dominated
model, with $\Omega _{\Lambda }=0.7$, $\Omega _{m}=0.3$, $\Omega
_{b}h^{2}=0.02$, $h=0.7$ and almost scale invariant spectrum $n=-2.9$. The
dotted curve (....) obtains when one changes to $\Omega _{m}=1$ and $\Omega
_{\Lambda }=0$ model. The dashed-triple-dotted curve (--$\cdot \cdot \cdot $--) is
for the  $\Lambda $-dominated model with a larger baryon density $\Omega
_{b}h^{2}=0.03$, while the dashed curve (-- -- --) changes this model by
adopting a magnetic spectral index of $n=-2.5$. We also show for qualitative
comparison (dashed-dotted curve -- $\cdot$ -- ), the temperature anisotropy in a
'standard'  $\Lambda $-CDM model, computed using CMBFAST
(Seljak \& Zaldarriaga 1996), with cosmological parameters as for the
first model described above.
These curves show the build up of power due to vortical perturbations
from tangled magnetic fields which survive Silk damping at high
$l \sim 1000-3000$. The eventual slow decline is due to the
damping by photon viscosity, although this decline
is only a mild decline as the magnetically sourced vortical mode is overdamped.
By contrast, in the absence of magnetic tangles there is a
sharp cut off due to Silk damping. }
\end{figure}

\begin{figure}
\begin{picture}(240,240)
\psfig{figure=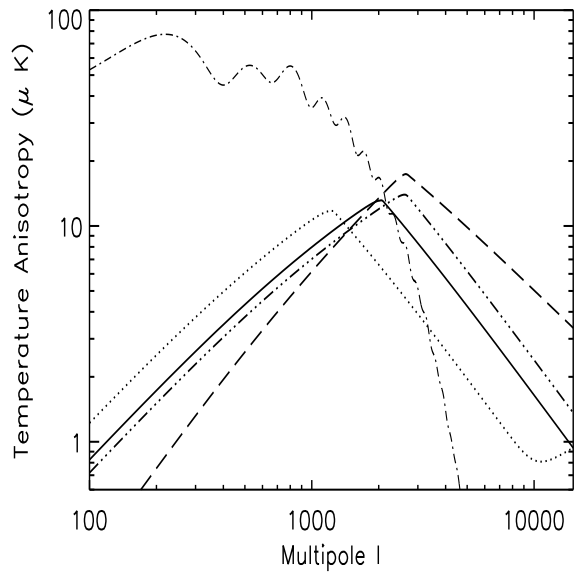,height=8.5cm,width=8.5cm,angle=0.0}
\end{picture}
\caption{$Log(\Delta T)$ versus $Log(l)$ predictions  up to very large $%
l\sim 10^{4}$; the line codings are same as in Figure 1.}
\end{figure}

\section{Discussion}

We re-examined the expected small-angular scale anisotropy induced by tangled
magnetic fields, for different cosmological parameters and spectral indices
and including a new regime at very small scales. We are motivated primarily
by several ongoing and future experiments, which probe this large- $l$
regime. Tangled magnetic fields give a distinctive contribution to the high-$%
l$ signal, since they create vortical perturbations that survive Silk
damping on much smaller scales than do compressional modes. Moreover, for
small-scale rotational perturbations, the damping due to the finite
thickness of the LSS is also milder than for compressional modes. It should
be emphasized that, by contrast, in the standard non-magnetic models the $%
C_{l}s$ have a sharp cut off for $l>k_{S}R_{\ast }$, due to Silk damping. As
we see from Figures 1 and 2, it is precisely here that the magnetically
induced signals begin to dominate. A scale-invariant spectrum of tangled
fields which redshifts to a present value $B_{0}=3\times 10^{-9}$ Gauss,
produces temperature anisotropies at the $10\mu K$ level between $l\sim
1000-3000$. Larger signals are produced in a $\Lambda $-dominated universe,
in a universe with larger baryon density, or if the spectral index of
magnetic tangles is steeper, $n>-3$. Note that the anisotropy in hot or cold
spots could be several times larger, because the non-linear dependence of $%
C_{l}$ on $M(k)$ will imply non-Gaussian statistics for the anisotropies.
Indeed, as we pointed out in Paper I , the $\Delta T(l)$ will obey a
modified form of the $\chi ^{2}$ distribution, the probability falling off
almost exponentially, rather than as a Gaussian.

We see that the predicted signal is quite sensitive to both the cosmology
and also the baryonic content (the value and any clumping of $\Omega _{B}$).
It is also necessary in future work to calculate the damping effects more
accurately, and also to take account of the effects of non-Gaussianity. The
latter point is especially important in searches which cover limited regions
of the sky, and the variance may not be the best statistic to compare with
observations. It is likely that the interpretation of the data at these
small angular scales will also be complicated by the need to understand the
contribution from discrete sources.

The other potentially important source of temperature anisotropies at these
small-angular scales is the Sunyaev-Zeldovich signal. However, it should  be
possible to isolate this signal by its frequency dependence. An important
means of distinguishing the effects of tangled magnetic fields is by looking
at the polarisation of the CMBR. It was shown by Seshadri and Subramanian
(2001) that the vorticity induced by tangled magnetic fields produces
significant and predominantly B-type polarization, at the micro-Kelvin
levels, and this will also be dominant at large $l$. These should therefore
be distinguishable from anisotropies produced by inflationary scalar and
tensor perturbations. We hope to revisit this issue in a future study.

Although we have concentrated on the vorticity induced anisotropy which is
dominant on small-angular scales, tangled magnetic fields also produce 
anisotropies on larger angular scale, dominated by tensor metric
perturbations induced by anisotropic magnetic stresses (Durrer, Ferreira \&
Kahniashvili 2000; Mack, Kahniashvili \& Kosowsky 2002). Using the formalism
described in these papers, we estimate a tensor contribution at small $l<100$
of $\sim 10.9\mu K(l/100)^{0.1}$ for the nearly scale invariant spectrum
with $n=-2.9$ and $\Delta T(l)\sim 4.9\mu K(l/100)^{0.5}$ for $n=-2.5$ and $%
B_{-9}\sim 3$. Since we have to add this power to the standard power
produced by inflationary scalar perturbations, in quadrature, a tangled field
with $B_{-9}\sim 3$ will produce of order a few to 10 percent
perturbation to the power in the standard CMBR anisotropy at small $l$s. So
if they are indeed detected at large $l$, below the Silk damping scale, and
if they have a power-law spectrum extending to large-scales, one will also
have to consider their effects seriously at large-angular scales, especially
in cosmological parameter estimation.

We also note that a magnetic field which redshifts to a present day value of
order a nano Gauss could impact significantly on the formation of galaxies
(Rees \& Reinhardt 1972, Wasserman 1978, SB98). Our results  therefore
provide a way of detecting the presence of small-scale magnetic
inhomogeneities at a level which affects the formation of galaxies and
clusters.

\section*{Acknowledgments}

KS thanks Simon White and Max-Plank Institut for Astrophysics, Garching, for
hospitality during the course of this work. We would like to thank P.
Ferreira, R. Crittenden, A. Lewis and A. Lasenby for discussions.



\begin{thebibliography}{99}
\bibitem{barrow}  Barrow J. D., Ferreira P. G., Silk J., 1997, Phys. Rev.
Lett., 78, 3610

\bibitem{beck}  Beck R., Brandenburg A., Moss D., Shukurov A. M., Sokoloff
D. D., 1996, Ann. Rev. Astron. Astrophys., 34, 155

\bibitem{BF00}  Blackman E. G., Field G. F., 2000, Ap.J., 534, 984 

\bibitem{B01}  Brandenburg A., 2001, Ap.J., 550, 824 

\bibitem{BS00}  Brandenburg A., Subramanian K., 2000, A\&A, 361, L33

\bibitem{BDS02}  Brandenburg A., Dobler W., Subramanian K., 2002, AN, 323, 99 

\bibitem{cat}  Cattaneo F., Vainshtein S. I., 1991, Ap.J., 376, L21

\bibitem{ch}  Cattaneo F., Hughes D. W., 1996, Phys. Rev. E, 54, 4532

\bibitem{capdur}  Caprini C., Durrer R., 2002, Phys. Rev. D, 65, 3517

\bibitem{clarke}  Clarke T. E., Kronberg P. P., Bohringer H., 2001, Ap.J.,
547, L111

\bibitem{dur}  Durrer R., Ferreira P. G., Kahniashvili T., 2000, Phys. Rev.
D, 61, 043001

\bibitem{rubgras}  Grasso D., Rubinstein H. R., 2001, Phys. Rep., 348, 161

\bibitem{GD}  Gruzinov A. V., Diamond P. H., 1994, Phys. Rev. Lett., 72, 1651

\bibitem{huwhita}  Hu W., White M., 1997a, Phys. Rev. D, 56, 596

\bibitem{huwhit}  Hu W., White M., 1997b, Ap.J., 479, 568

\bibitem{husug}  Hu W., Sugiyama N., 1995, Ap.J., 444, 489

\bibitem{jko}  Jedamzik K., Katalinic V., Olinto A., 1998, Phys. Rev. D, 57,
3264

\bibitem{kor}  Kim E. J., Olinto A. V., Rosner R., 1996, Ap.J., 468, 28 

\bibitem{primhyp}  Kulsrud R. M., IAU Symp. 140: {\it Galactic and
Extragalactic Magnetic Fields}, Reidel, Dordrecht, (1990), p527

\bibitem{kul99}  Kulsrud R. M., 1999, Ann. Rev. Astron.  Astrophys., 37, 37

\bibitem{ka}  Kulsrud R. M., Anderson S. W., 1992, Ap.J., 396, 606

\bibitem{klee}  Kleeorin N., Moss D., Rogachevskii I., Sokoloff D., 2000,
A\&A, 361, L5

\bibitem{kron}  Kronberg P. P., 1994, Rep. Prog. Phys., 57, 325

\bibitem{mack}  Mack A., Kashniashvili T., Kosowsky A., 2002, Phys. Rev. D
(in Press), astro-ph/0105504

\bibitem{ow}  Oren A. L., Wolfe A. M., 1995, Ap.J., 445, 624

\bibitem{reesrein}  Rees M. J., Reinhardt M., 1972, A\&A, 19, 189

\bibitem{rss}  Ruzmaikin A. A., Shukurov A. M., Sokoloff D. D., 1988, {\it %
Magnetic Fields of Galaxies}, Kluwer, Dordrecht (1988)

\bibitem{selzal} Seljak U., Zaldarriaga M., 1996, Ap.J., 469, 437

\bibitem{seshsub}  Seshadri T. R., Subramanian K., 2001, Phys. Rev. Lett.,
87, 101301

\bibitem{ravi}  Subrahmanyan R., Kesteven M. J., Ekers R. D., Sinclair M.,
Silk J., 2000, Mon. Not. R. astron. Soc, 315, 808

\bibitem{silk}  Silk J., Ap.J., 1968, 151, 431

\bibitem{suba}  Subramanian K., 1998, Mon. Not. R. astron. Soc, 294, 718

\bibitem{ks99}  Subramanian K., 1999, Phys. Rev. Lett., 83, 2957

\bibitem{sb98}  Subramanian K., Barrow J. D., 1998, Phys. Rev. D, 58, 083502
(SB98)

\bibitem{ksjd2}  Subramanian K., Barrow J. D., 1998, Phys. Rev. Lett., 81,
3575 (Paper I)

\bibitem{tw}  Turner M. S., Widrow L. M., 1998, Phys. Rev. D, 30, 2743

\bibitem{ratra}  Ratra B., 1992, Ap.J. Lett., 391, L1

\bibitem{wasser}  Wasserman I., Ap.J., 1978, 224, 337
\end{thebibliography}
\end{document}